\documentclass[12pt]{scrartcl}
\areaset{158mm}{238mm}
\usepackage{setspace}
\usepackage{amsmath}
\usepackage{mathtools}

\usepackage{amssymb}
\usepackage[mathscr]{eucal}
\usepackage[unicode=true,linktoc=all]{hyperref}

\usepackage{cite}
\usepackage{bm}
\usepackage{braket}
\usepackage{slashed}

\newcommand\sign{\mathop{\mathrm{sign}}\nolimits}

\numberwithin{equation}{section}

\makeatletter
\DeclareFontFamily{OMX}{MnSymbolE}{}
\DeclareSymbolFont{MnLargeSymbols}{OMX}{MnSymbolE}{m}{n}
\SetSymbolFont{MnLargeSymbols}{bold}{OMX}{MnSymbolE}{b}{n}
\DeclareFontShape{OMX}{MnSymbolE}{m}{n}{
    <-6>  MnSymbolE5
   <6-7>  MnSymbolE6
   <7-8>  MnSymbolE7
   <8-9>  MnSymbolE8
   <9-10> MnSymbolE9
  <10-12> MnSymbolE10
  <12->   MnSymbolE12
}{}
\DeclareFontShape{OMX}{MnSymbolE}{b}{n}{
    <-6>  MnSymbolE-Bold5
   <6-7>  MnSymbolE-Bold6
   <7-8>  MnSymbolE-Bold7
   <8-9>  MnSymbolE-Bold8
   <9-10> MnSymbolE-Bold9
  <10-12> MnSymbolE-Bold10
  <12->   MnSymbolE-Bold12
}{}

\let\llangle\@undefined
\let\rrangle\@undefined
\DeclareMathDelimiter{\llangle}{\mathopen}%
                     {MnLargeSymbols}{'164}{MnLargeSymbols}{'164}
\DeclareMathDelimiter{\rrangle}{\mathclose}%
                     {MnLargeSymbols}{'171}{MnLargeSymbols}{'171}
\makeatother

\title{\vspace{2cm} Worldsheet Geometries of Ambitwistor String\vspace{1.5 cm}}
\author{Kantaro Ohmori
\\[0.5cm]
	\large\slshape Department of Physics, the University of Tokyo,\\[-0.2cm] 
	\large\slshape Hongo, Bunkyo-ku, Tokyo 133-0022, Japan \\[-0.05cm]
}
\date{}
\begin{document}
\vspace{4cm}
\maketitle

\vspace{-11cm}
\hfill \\[-.5cm]

\hfill  UT-15-09\\[-.5cm]

\vspace{11cm}
\paragraph{\hspace{.9cm}\large{Abstract}}
\vspace{-.1cm}
\begin{abstract}
	Mason and Skinner proposed the ambitwistor string theory which directly reproduces the formulas for the amplitudes of massless particles proposed by Cachazo, He and Yuan.
	In this paper we discuss geometries of the moduli space of worldsheets associated to the bosonic or the RNS ambitwistor string.
	Further, we investigate the factorization properties of the amplitudes when an internal momentum is near on-shell in the abstract CFT language.
	Along the way, we propose the existence of the ambitwistor strings with three or four fermionic worldsheet currents.
\end{abstract}
\thispagestyle{empty}
\newpage
\setcounter{tocdepth}{2}
\tableofcontents

\section{Introduction}
In \cite{Mason:2013sva}, the ambitwistor string, which reproduces the CHY formulas \cite{Cachazo:2013hca,Cachazo:2013iea} for Einstein gravity, Yang-Mills, and scalars with cubic interactions was proposed as a generalization of the twistor string proposed in \cite{Witten:2003nn}.
The bosonic and non-compact part of the worldsheet action of the ambitwistor string is
\begin{align}
	S=\frac1{2\pi}\int \left(\eta^{\mu\nu}P_\mu\bar{\partial}X_\nu -\frac12 e\eta^{\mu\nu} P_\mu P_\nu\right),
\end{align}
where $X_\mu$ are worldsheet scalars valued in the complexified space-time $\mathbb{C}^D$
and $P_\mu$ are worldsheet one-forms which are conjugate to $X^\mu$.
The Lagrange multiplier $e$ enforces $P^2=0$, and the gauge symmetry
\begin{align}
	\delta e=\bar{\partial}\alpha,\qquad \delta X_\mu=\alpha P_\mu
\end{align}
generated by the spin 2 current $\tilde T^\text{m}:=-\frac12P^2$ reduces the target space down to the ambitwistor space $\mathbb{A}$. More precisely, the action $S$ is regarded as the integration of the pull back of the holomorphic contact structure on the projective ambitwistor space $\mathbb{PA}$.

The striking property of the ambitwistor string is that amplitude $\mathcal{A}_\text{usual}$ is always localized at isolated points on the moduli space, which are the solutions of the scattering equations.
The scattering equations, for the tree level $n$-points scattering with external momenta $k_a,\,(a=1,\cdots,n)$, are
\begin{align}
	\sum_{b\neq a}\frac{k_a\cdot k_b}{\sigma_a-\sigma_b}=0
\end{align}
for $\sigma_a\in\mathbb{C}$ and their solutions determines $(n-3)!$ points of the moduli $\mathcal{M}_{0,n}$ of $n$-punctured spheres.
Scattering equations have already appeared in \cite{Gross:1987kza} as a saddle-point equation in the semi-classical analysis of the usual string, and are recently actively investigated starting from \cite{Cachazo:2013gna} and their following papers.

The localization is caused by the integrated form of the vertex operators which looks like
\begin{align}
	\int_\Sigma \bar{\delta}(k\cdot P)V^\text{m},
	\label{eq:intverintro}
\end{align}
where $V^\text{m}$ is a matter primary transforming as a $(1,0)$-form on the worldsheet $\Sigma$, containing the kinematical factor $e^{\mathrm{i}k\cdot X}$.
The delta-from $\bar{\delta}(k\cdot P):=\bar{\partial}\frac{1}{k\cdot P}$
is needed to integrate over the worldsheet $\Sigma$, and is suggested by the Penrose transformation of corresponding excitations on the space-time explained in \cite{Mason:2013sva}.

However, how should the insertion of the delta-forms be understood from the purely worldsheet point of view?
The first objective of this paper is to clarify how the amplitudes should be defined and where the delta-forms come from.
As stated in \cite{Adamo:2013tsa}, the delta-form should come from the integration in terms of the ``moduli'' associated to the Lagrange multiplier $e$, but no full geometrical treatment is found in the literature yet.

The ambitwistor string is not a usual string, because the worldsheet theory has two chiral spin 2 gauge symmetries, generated by the stress energy tensor $T$ and an additional weight 2 primary $\tilde{T}^\text{m}=-\frac12P^2$, while the usual closed string theory has chiral and anti-chiral conformal symmetries.
Therefore, we should check step-by-step whether the logic we have used in the case of usual string can be applied to the ambitwistor string.

In the usual string theory, the on-shell amplitudes are defined as the integration over the moduli space $\mathcal{M}$ of punctured Riemann surfaces:
\begin{align}
	\mathcal{A}_\text{usual}=\int_{\mathcal{M}} F
\end{align}
where $F$ is a top-form on $\mathcal{M}$ defined by the worldsheet correlator.
The required features, namely decoupling of BRST-exact states and factorization of amplitudes when internal momenta go on-shell, can be proven via this definition.

In the ambitwistor string, things do not go in exactly the same way.
The worldsheet $\Sigma$ couples to two backgrounds, the metric $g$ and the Lagrange multiplier Beltrami differential $e\in \Omega^{(0,1)}(T\Sigma)$.
Therefore, the correlators on $\Sigma$ define a form $F$ on the cotangent bundle $T^*\mathcal{M}$ rather than that on $\mathcal{M}$ itself. 
Moreover, because the worldsheet theory do not have anti-chiral part, the defined form $F$ is a holomorphic top-form when all the vertex insertions are physical.
Thus we cannot integrate $F$ over $T^*\mathcal{M}$ nor restrict $F$ on the zero section $\mathcal{M}\subset T^*\mathcal{M}$ to define amplitudes since $F|_\mathcal{M}=0$.
We need a non-trivial middle dimensional cycle $\Gamma$ in $T^*\mathcal{M}$ to integrate $F$ over.

In this paper, we propose to construct such a cycle $\Gamma$ by the Morse flow, explained in the section \ref{sec:bos}. The amplitude $\mathcal{A}_\text{ambitwistor}$ are defined as 
\begin{align}
	\mathcal{A}_\text{ambitwistor} =\int_{\Gamma\subset T^*\mathcal{M}} F.
\end{align}
While the cycle $\Gamma$ is complicated, we will find that the localization holds with respect to the BRST-operator $Q$ and the integration reduces to a sum over points in $\mathcal{M}$ which are indeed the solutions of the scattering equations.
In the end, the result reproduces the amplitudes obtained by using the formal integrated vertices \eqref{eq:intverintro}.

The second goal of this paper is to deduce the factorization properties and the propagators of ambitwistor string. 
In usual closed string, the amplitude evaluated on cylinders gives the propagator \cite{Witten:2012bg}
\begin{align}
	\frac{b_0\tilde{b}_0\delta(L_0-\overline{L}_0)}{L_0}.
\end{align}
For the ambitwistor string, this is not the desired form because $L_0$ do not contain the kinematical factor $k^2$,
therefore something else should happen.

We will find that there exist solutions of scattering equations on the boundary of the moduli space when the corresponding internal momentum gets on-shell, as is directly proven in \cite{Cachazo:2013gna} at the tree level.
The degeneration parameter $q_*$ of the solutions of the scattering equations look
\begin{align}
	q_* \propto k^2 +\mathcal{O}((k^2)^2).
\end{align}
Then, the singular part of the amplitude of the bosonic ambitwistor string is captured by the operator
\begin{align}
	\frac{q_*^{-2}\delta(L_0-2)+q_*^{-1}\delta(L_0-1)+\delta(L_0)}{k^2}.
\end{align}
The first two terms in the numerator are unphysical since they give cubic and quadratic poles of $k^2$.
However, in the case of type II ambitwistor string, we will find that such terms are eliminated by the GSO projection.
Moreover, there is no contribution to the (double) leading trace part of 
the bosonic or ``heterotic'' ambitwistor string which contain the $\phi^3$ theory or the Yang-Mills theory, respectively.
Thus there is no contradiction with the literature.

As a by-product, we will find there exists ambitwistor string with three or four fermionic currents.
Their all-NS sector reproduces CHY representations for the DBI theory and a special kind of Galileon theory.
We will quickly mention on them in section~\ref{sec:conc}.

In this paper we focus on RNS formalism of ambitwistor string 
and will not touch the pure spinor version \cite{Berkovits:2013xba} of that.
There are also variants of the original ambitwistor string of \cite{Mason:2013sva}
which have different the bosonic part \cite{Geyer:2014fka,Geyer:2014lca,Adamo:2014wea,Adamo:2015fwa}.
It will be interesting to consider the factorization properties of those theories.

\section{Amplitudes from a bosonic chiral CFT}
\label{sec:bos}

\subsection{Algebras and the BRST charge}
To construct the bosonic ambitwistor string, we need a triple
\begin{align}
	(\mathcal{T},T^\text{m},\tilde{T}^\text{m}),
\end{align}
where $\mathcal{T}$ is a 2d CFT, $T^\text{m}$ is the holomoriphic part of its stress energy tensor, 
and $\tilde{T}^\text{m}$ is another primary operator of weight 2.
In this paper, we assume that $\mathcal{T}$ should be decomposed into $\mathcal{T}_\text{flat}\oplus\mathcal{T}_\text{int}$ where $\mathcal{T}_\text{flat}$ is the flat $XP$ CFTs representing the projective ambitwistor space $\mathbb{PA}$ of the $D\ge 2$ dimensional flat space and $\mathcal{T}_\text{int}$ is a internal compact CFT.
If only the tree level amplitudes are considered, the modular invariance of $\mathcal{T}$ is not needed.
The central charge $c^\text{m}$ of $\mathcal{T}$ will be constrained later.

A known meaningful internal system $\mathcal{T}_\text{int}$ is the sum of two current algebras $\hat{\mathfrak{g}}\oplus\hat{\tilde{\mathfrak{g}}}$.
The double leading trace part for those is known to give the CHY formula given in \cite{Cachazo:2013iea} for scalars $\phi_{a\tilde{a}}$ which have adjoint indices for each current algebra with the cubic interaction $f_{abc}\tilde{f}_{\tilde a\tilde b \tilde c}\phi^{a\tilde a}\phi^{b\tilde b}\phi^{c\tilde c}$ as shown in \cite{Mason:2013sva}.

We also assume that the spin 2 current $\tilde{T}$ is equal to the operator $-\frac12 P^2$ of 
$\mathcal{T}_\text{flat}$.
The OPE between those currents are
\begin{align}
	T^\text{m}(z)T^\text{m}(0)&\sim \frac{c^\text{m}}{2z^4}+\frac{2}{z^2}T(0)+\frac{1}{z}\partial T(0), \nonumber\\
	T^\text{m}(z)\tilde{T}^\text{m}(0) &\sim \frac{2}{z^2}\tilde{T}^\text{m}(0)+\frac{1}{z}\partial \tilde{T}^\text{m}(0), \nonumber\\
	\tilde{T}^\text{m}(z)\tilde{T}^\text{m}(0) &\sim 0.
	\label{eq:TtilTOPE}
\end{align}

To gauge $\tilde{T}^\text{m}$, we couple this current to the background field $e=e_{\bar{z}}^z$ which transforms as a Beltrami differential on the Riemann surface $\Sigma$ and integrate in the combination $e\tilde{T}$ over $\Sigma$.
In other words, we put the additional term
\begin{align}
	S_{e\tilde{T}^\text{m}}= \frac1{2\pi}\int_\Sigma e\tilde{T}^\text{m} =: (e,\tilde{T}^\text{m}).
\end{align}
in the action. We will use this pairing $(\,,\,)$ of a Beltrami differential and a quadratic differential in the rest of this paper. Note that $e$ is independent of the Beltrami differential $\mu$ which is defined by the variations of the metric.

To quantize the system,
we add a pair of ghost systems, namely $bc$ and $\tilde{b}\tilde{c}$ systems. Both are usual \emph{holomorphic} ghost systems thus both $b$ and $\tilde{b}$ have holomoriphic weight 2 and their holomorphic stress-energy tensors are $T_{bc}=-(\partial b)c-2b\partial c$ and $T_{\tilde{b}\tilde{c}}=-(\partial \tilde{b})\tilde{c}-2\tilde{b}\partial \tilde{c}$.
Then we can define the BRST weight 1 primary.
\begin{align}
	j_B= cT^\text{m}+cT_{\tilde{b}\tilde{c}}+\frac12 cT_{bc}+\tilde{c}\tilde{T}^\text{m}+\frac32\partial^2 c.
\end{align}
The last total derivative term just ensures $j_B$ to be a current.
The residue of the OPE between two of this current is
\begin{align}
	\mathrm{Res}\;j_B(z)j_B(0)= \frac{52-c^\text{m}}{12}c\partial^3c+\partial(c\tilde{c}\tilde{T}^\text{m}),
\end{align}
therefore the BRST charge $Q=\oint \frac{\mathrm{d}z}{2\pi\mathrm{i}}j_B(z)$ is nilpotent when $c^\text{m}=52$.
In the following we set the central charge $c^\text{m}$ to be $52$.

In the usual string theory, the important property of the ghost system was
\begin{align}
	Q\cdot b = T.
\end{align}
This equation also holds in the present context, but for $\tilde{b}\tilde{c}$ system we have
\begin{align}
	Q\cdot \tilde{b} = \tilde{T}^\text{m} - (\partial\tilde{b})c-2\tilde{b}\partial c.
\end{align}
We should recognize the remnant term $\tilde{T}^{\text{gh}} := - (\partial\tilde{b})c-2\tilde{b}\partial c$ as the ghost contribution to the total distinguished spin 2 current $\tilde{T} := \tilde{T}^\text{m}+\tilde{T}^{\text{gh}}$ which couples to the Beltrami differential background $e$.
Hence, we have the term
\begin{align}
	S_{e\tilde{T}}= (e,\tilde{T}).
\end{align}
in the total action on the worldsheet.
\footnote{It may seem more natural to define $j_B$ as 
$j_B= cT^\text{m}+\frac12 cT_{\tilde{b}\tilde{c}}+\frac12 cT_{bc}+\tilde{c}\tilde{T}^\text{m}+\frac12\tilde{c}\tilde{T}^\text{gh}$, but both expression are essentially the same since $cT_{\tilde{b}\tilde{c}} -\tilde{c}\tilde{T}^\text{gh}$ is a total derivative.}

\subsection{Vertex operators}
In this paper, we only consider vertex operators $\mathcal{V}$ which vanish by the action of
$b_n$ and $\tilde{b}_n$ for $n\ge 0$.
We call operators of this type standard vertex operators.
Equivalently, $\mathcal{V}$ does not contain any derivatives of $c$ and $\tilde{c}$.
If a standard vertex $\mathcal{V}$ has the ghost number $2$ and $Q$-closed, we call it a physical vertex, and such $\mathcal{V}$ takes the form of
\begin{align}
	\mathcal{V}=c\tilde{c}V^\text{m}
	\label{eq:physV}
\end{align}
where $V^\text{m}$ is a vertex operator of the matter system $\mathcal{T}$.
$Q$-closedness of $\mathcal{V}$ also requires that $\mathcal{V}$ vanishes when acted by $L_n=\{Q,b_n\}$ and $\tilde{L}_n:= \oint\frac{\mathrm{d}z}{2\pi\mathrm{i}}z^{n+1}\tilde{T}(z)=\{Q,\tilde{b}_n\}$ for $n\ge0$,
thus $V^m$ should be a primary of weight 2 and should be annihilated by $\tilde{L}_n$ for $n\ge0$.
We are always able to take a representative of $Q$-cohomology of this form, which is proven in Appendix. \ref{sec:noghost}.

For the flat space bosonic ambitwistor string, $V^\text{m}$ is taken to be
\begin{align}
	V^\text{m}=\epsilon^{\mu\nu}P_\mu P_\nu e^{\mathrm{i}k\cdot X},
	\label{eq:flatvert}
\end{align}
then the first condition requires $\epsilon^{\mu\nu}k_\mu=0$ and the second condition does $k^2=0$.
Note that the kinematic factor $e^{\mathrm{i}k\cdot X}$ do not have conformal weight
as opposed to the usual string case,
therefore the physical operators are more constrained and the number of them is finite.

If we include two current algebras $\hat{\mathfrak{g}}\oplus\hat{\tilde{\mathfrak{g}}}$,
there are also physical operators whose matter part is
\begin{align}
	V^\text{m}= j_a \tilde{\jmath}_{\tilde{a}} e^{\mathrm{i}k\cdot X},
\end{align}
where $j_{a}$ and $\tilde{\jmath}_{\tilde{a}}$ are weight 1 current operators.
In \cite{Mason:2013sva} it was found that the double leading trace term of $j$ and $\tilde{\jmath}$ correlators gives the CHY formulas for scalars $\phi_{a\tilde{a}}$ with the interaction $f_{abc}\tilde{f}_{\tilde{a}\tilde{b}\tilde{c}}\phi^{a\tilde a}\phi^{b\tilde b}\phi^{c\tilde c}$.

To consider the decoupling of $Q$-exact vertices, we also assume that when an operator $\mathcal{V}$ of the form of \eqref{eq:physV} is $Q$-exact, there exists a standard vertex operator $\mathcal{W}$ which satisfies $Q\cdot \mathcal{W}=\mathcal{V}$. This should be shown as in the appendix of \cite{Witten:2012bh}, but we just postulate it here.
\footnote{
	The proof of \cite{Witten:2012bh} is based on the proof of the no-ghost theorem. We address how to modify the proof of the no-ghost theorem for the ambitwistor case in the appendix \ref{sec:noghost}, and the similar modification is expected to be available for the statement here.
}

\subsection{Holomorphic top-form on the cotangent bundle of the moduli}
To define the amplitude of the ambitwistor string, we need a integration measure on a suitable moduli space.
We will see that the above BRST structure defines a holomorphic top-form on the cotangent bundle of the moduli space,
which cannot be integrated over the whole cotangent bundle.
Therefore, we should define a integration cycle to integrate that holomorphic top-form, which is treated in section~\ref{sec:Intcycle}.
Here we will adopt the ``extended formalism'' of \cite{Witten:2012bh}.

First, we obtain a form on the space $T^*\mathscr{J}$,
where $\mathscr{J}=\{\text{metric on $\Sigma$}\}/(\text{Weyl transf.})$ is  the space of conformal structures on the considered Riemann surface and $T^*\mathscr{J}$ is its holomorphic cotangent bundle whose coordinates can be locally represented by a pair of a metric and a Beltrami differentials $(g,e)$.
To define the form, we consider additional fermionic fields $\delta g ,\delta e$ both transforming as Beltrami differentials, and extend the definition of $Q$ so that it acts on $\mu$ and $e$ as
\begin{align}
	[Q,g]=\delta g,\qquad[Q,e]=\delta e.
\end{align}
To keep the action $Q$-invariant, we add the extra term
\begin{align}
	S_\text{ext}=(\delta \mu,b)+(\delta e,\tilde{b}).
\end{align}
We recognize $\delta g(z),\delta e(z)$ as differential 1-forms on $T^*\mathscr{J}$ associated to local coordinates $(g,e)$.
Note that the differential 1-forms $\delta \mu(z) :=g^{z\bar{z}}\delta g_{\bar z \bar z}$ span holomorphic differential 1-forms together with $\delta e(z)$.

To define a nontrivial form on $T^*\mathscr{J}$, choose fixed positions $z_i$ on the Rieman surface where the $i$th ($i=1\cdots n$) standard vertex $\mathcal{V}_i$ sits. 
Then we define a quantity
\begin{align}
	F(\Omega;g,e|\delta \mu,\delta e)
	:=\left\langle e^{-(\delta\mu,b)} e^{-(e,\tilde{T})-(\delta e,\tilde{b})}  \Omega \right\rangle_{g},
	\label{eq:Fdef}
\end{align}
where $\langle \rangle_{g}$ is the correlator of $\mathcal{T}$ and the ghost systems with a metric background $g$ and operator insertions $\Omega=\prod_i^n \mathcal{V}_i(z_i)$.
$F$ is the partition function on the worldsheet with vertex operator insertions and with additional actions $S_\text{ext}$ and $S_{e\tilde{T}}$.
\footnote{ Here we used a renormalization scheme which satisfy $\frac\delta{\delta g_{\bar z\bar z}}\langle\mathcal{O}\rangle_g=\langle -T \mathcal{O}\rangle_g$ where $\mathcal{O}$ is a scalar operator and $T$ is inserted in the leftmost in the bracket. Then \eqref{eq:Fdef} is invariant under the extended $Q$ up to $Q\cdot \Omega$. 
The more precise description of the regularization is in subsection~\ref{sec:remark}.}
Identifying $\delta \mu$ and $\delta e$ with holomorphic differential forms, $F(\{\mathcal{V}_i\};\mu,e|\delta\mu,\delta e)$ defines a holomorphic differential form over a patch around $g$ and extends globally over $T^*\mathscr{J}$.
Counting the ghost numbers, the degree of the holomorphic form $F$ is $6g-6+N_\text{gh}$ where $N_\text{gh}$ is the total ghost number (the sum of $bc$ and $\tilde{b}\tilde{c}$ ghost number) of $\Omega$. 
Moreover, since the evaluation of $\langle \rangle_{g}$ is a BRST-invariant operation and $Q$ is equivalent to the exterior derivative $\mathrm{d}$ on $T^*\mathscr{J}$ when it acts on the function only of $g,e,\delta \mu,\delta e$, we get
\begin{align}
	\mathrm{d}F(\Omega)+F(Q\cdot \Omega)=0.
	\label{eq:dF}
\end{align}
In particular, if all the standard vertex operators $\mathcal{V}_i$ are $Q$-closed, $F$ is a closed form.

We want to show the form $F$ over $T^*\mathscr{J}$ is actually a pullback of a form over the holomorphic cotangent bundle $T^*\mathcal{M}$ of the moduli space $\mathcal{M}$ of Riemann surfaces with $n$ punctures.
The space $T^*\mathcal{M}$  can be obtained by a quotient
\begin{align}
	T^*\mathcal{M}=T^*\mathscr{J}/\mathcal{D}
\end{align}
where the infinite dimensional group $\mathcal{D}$ is generated by a pair of vectors $(v=v^z,\alpha=\alpha^z)$ which satisfy $v(z_i)=\alpha(z_i)=0$.
$v$ is the parameter of diffeomorphisms and $\alpha$ is that of the gauge transformations generated by $\tilde{T}$.

To show $F$ is obtained by pulling back a form over $T^*\mathcal{M}$,
we should ensure that $F$ is invariant under $\mathcal{D}$ and $F$ vanishes when contracted with a vector field $\mathcal{X}_{(v,\alpha)}$ which is generated by the action of an element $(v,\alpha)$ of $\mathcal{D}$.
The invariance of $F$ under $\mathcal{D}$ is manifest by the imposed symmetry.
$\mathcal{X}=\mathcal{X}_{(v,\alpha)}$ acts on the coordinates $(g,e)$ as
\begin{align}
	\mathcal{X}\cdot(g_{\bar{z}\bar{z}},e) = (g_{\bar zz}\bar\partial_{\bar z}v^z,\bar{\partial}\alpha + v\partial e-(\partial v) e).
\end{align}
Then the contraction with $F$ is calculated by
\begin{align}
	i(\mathcal{X})F &=\left(g^{z\bar z}(\mathcal{X}\cdot g_{\bar z \bar z})\frac{\delta}{\delta \,\delta\mu}+(\mathcal{X}\cdot e)\frac{\delta}{\delta \, \delta e}\right)F(\mu,e|\delta \mu,\delta e)\nonumber\\
	      &= - \left\langle e^{-(\delta \mu,b)}
	\left((\bar{\partial}v,b)+(\mathcal{X}\cdot e,\tilde{b})\right)e^{-(e,\tilde{T})-(\delta e,\tilde{b})}\Omega\right\rangle
	\nonumber\\
	&= - \frac1{2\pi}\left\langle e^{-(\delta \mu,b)}
	\int_\Sigma\left( v(-\bar\partial b-e\partial \tilde{b}+2\partial(e\tilde{b}))-\alpha\bar\partial \tilde{b}\right)e^{-(e,\tilde{T})-(\delta e,\tilde{b})}\Omega\right\rangle
	\label{eq:iXF}.
\end{align}
We did some integrations-by-parts from the second line to the last line.
The equations of motion for $c$ and $\tilde{c}$ are
\begin{align}
	-\bar\partial b-e\partial \tilde{b}+2\partial(e\tilde{b})&=0 \nonumber\\
	-\bar\partial \tilde{b}&=0,
\end{align}
concluding the vanishing of \eqref{eq:iXF}.
The contact terms between the equations of motion and vertices $\Omega=\prod c\tilde{c}V^\text{m}(z_i)$ are also prohibited because of the condition $v(z_i)=\alpha(z_i)=0$.
This is why we concentrate on standard vertex operators.

Now we are able to regard $F$ as a holomorphic $6g-6+N_\text{gh}$ form on $T^*\mathcal{M}$.
If all vertices $\mathcal{V}_i$ are physical, $N_\text{gh}=2n$ and $F$ is a holomorphic top-form
since $T^*\mathcal{M}$ has complex dimension $6g-6+2n$.
It is convenient to introduce a local coordinate $(t_i,a_i)$, $i=1,\cdots,\mathrm{dim}_\mathbb{C}\mathcal{M}$ of $T^*\mathcal{M}$ and 
choose representatives $g=g(t)$ and $e=e(t,a)$ of the quotient classes of $T^*\mathscr{J}/\mathcal{D}$.
What we have shown is that $F$ is independent of such choices.
We take the coordinates $a_i$ so that  $e$ is linear in $a$'s as $e=\sum_i a_ie_i$.
It is convenient to rewrite $F$ of \eqref{eq:Fdef} when restricted to the local patch $(t,a)$ around $g_0=g(t=0)$ in $\mathcal{M}$ as
\begin{align}
	F&=\left\langle e^{-\{Q,(\mu(t),b)\}}e^{-\{Q,(e(a),\tilde{b})\}}\Omega\right\rangle_{g_0}\nonumber\\
	 &=\left\langle e^{-(\mathrm{d} \mu(t),b)-(\mu(t),T)}e^{-(\mathrm{d} e(a),\tilde{b})-(e(a),\tilde{T})}\Omega\right\rangle_{g_0}
	\label{eq:Fdef2}
\end{align}
where $\mu(t)$ is the solution of $\partial_{t_i}\mu(t)=g^{z\bar z}(t)\partial_{t_i}g_{\bar z\bar z}(t)=:\mu_i(t)$,
and $\mathrm{d}$ is the exterior derivative on the finite dimensional space $T^*\mathcal{M}$.
Expanding the exponential factors in the definition \eqref{eq:Fdef} of $F$ including $b$ and $\tilde{b}$, we get ghost insertions as in the case of the usual string theory.

If one of $\mathcal{V}_i$ is $Q$-exact with $\mathcal{V}=Q\cdot \mathcal{W}$ and $\mathcal{W}$ is standard, we immediately conclude the $F$ is a total derivative by \eqref{eq:dF}, which indicates the decoupling of BRST-exact states.

The next step is to integrate $F$, but we should find a adequate middle cycle in $T^*\mathcal{M}$ over which $F$ is integrated. An immediate candidate is the zero section $\mathcal{M}$ of $T^*\mathcal{M}$, but $F$, which is proportional to $\mathrm{d}a_i$, becomes $0$ when restricted to where $a_i=0$, therefore we should do something nontrivial.

\subsection{Integration cycle and localization}
\label{sec:Intcycle}
To define a middle dimensional cycle in $T^*\mathcal{M}$, we propose to use the Morse theory in the manner of \cite{Witten:2010cx}.
As we will see, only the existence of the middle dimensional steep decay direction of $F$ from its critical points, which is ensured by the Morse theory, is important and concrete shape of the cycle will not play a role. 
We give a brief prescription of the definition of the cycle for completeness.

We use $h:=\Re \mathcal{I},$ with $\mathcal{I}=-(\mu(t),T)-(e(t,a),\tilde{T})$ as the Morse function on $T^*\mathcal{M}$ to define the integration cycle.
Then the critical points $p^m_*=(t^m_*,a^m_*)$ of $h$ satisfy
\begin{align}
	(e_j(t^m_*),\tilde{T})|_{g(t^m_*)}&=0, \label{eq:crit}\\
	\partial_{t_i}(e(a^m_*,t^m_*),\tilde{T})&=(\mu_i(t^m_*),T).
\end{align}
These equations mean the operators on the left hand sides and the right hand sides are equal when evaluated as additional insertions of the correlator $F$.
\footnote{For simplicity, we are ignoring the subtle ordering problems which will be clarified in section \ref{sec:remark}.}
We will see that the first equation which determines points $t^m_*$ in the moduli space $\mathcal{M}$ actually becomes the scattering equations.
Then the second equation which is linear in terms of $a_*^m$ fixes exactly one point in the fiber $T_{t^m}\mathcal{M}$. (We assume that $h$ is Morse, or the second equation is full rank.)

The Morse downward equation 
\begin{align}
	\frac{\mathrm{d}t_i}{\mathrm{d}\tau}=-\frac{\partial\bar{\mathcal{I}}}{\partial \bar{t}_i},\;\;
	\frac{\mathrm{d}a_j}{\mathrm{d}\tau}=-\frac{\partial\bar{\mathcal{I}}}{\partial \bar{a}_j}
	\label{eq:floweq}
\end{align}
defines the flows $u:\tau\mapsto (t(\tau),a(\tau))$ called the Stokes rays,
and the Lefschetz thimble $\mathcal{J}_m$ associated to the critical point $p_*^m$ is defined by 
\begin{align}
	\mathcal{J}_m:=\bigcup_{\substack{u:\text{Stokes ray}\\u(-\infty)=p^m_*}}\{u(\mathbb{R})\}\subset T^*\mathcal{M},
\end{align}
which is middle dimensional cycle in $T^*\mathcal{M}$ because the Morse function $h$ is the real part of a holomorphic function. Stokes rays generically do not connect distinct critical points since the flow preserves $\Im \mathcal{I}$, and we assume that.
We choose a sum of all the Lefschetz thimbles as the integration cycle $\Gamma$:
\footnote{ The integration can be defined with any cycles of the form $\sum n_m\mathcal{J}_m$ for some $n_m\in \mathbb{Z}$, though we choose $n_m=1$ for all $m$ just by hand to get the same result in the literature. A possible alternative is to set $n_m$ to be the intersection number of the Morse upward flow starting from $p_*^m$ and the zero section $\mathcal{M}\subset T^*\mathcal{M}$ imitating \cite{Witten:2010cx}, though we are not able to show this leads to the correct answer $n_m=1$. As will be mentioned in \ref{sec:loop}, how to choose $n_m$ is not clear for loop amplitudes.}
\begin{align}
	\Gamma := \sum_m  \mathcal{J}_m.
\end{align}
The amplitudes $\mathcal{A}_\Omega$ associated to physical insertions $\Omega$ are now defined as
\begin{align}
	\mathcal{A}_{\Omega}:= \int_{\Gamma}F(\Omega).
	\label{eq:defA}
\end{align}
Actually this definition does not work with $g\ge1$ since $F$ itself diverges because of the zero modes of $P^\mu$.
We postpone to deal with the loop amplitudes until subsection~\ref{sec:loop} and concentrate on the case of $g=0$ in this and the next subsection.

If $\Omega$ is a BRST-exact operator $\Omega=Q\cdot \Omega'$, the amplitude vanishes because of \eqref{eq:dF}:
\begin{align}
	\mathcal{A}_{Q\cdot \Omega'}=-\int_{\Gamma}\mathrm{d}F(\Omega')=0.
\end{align}

Finally, we find a drastic simplification of the amplitude $\mathcal{A}$.
Note that $F$ \eqref{eq:Fdef2} is defined by exponential of $Q$-exact quantities, which signals localization of the integration.
As usual, we scale the $Q$-exact exponential factors by a parameter $A$ and call it $F_A$:
\begin{align}
	F_A(\Omega):=
	\left\langle e^{-A \{Q,(\mu(t),b)\}}e^{-A \{Q,(e(t,a),\tilde{b})\}}\Omega\right\rangle_{g_0},
\end{align}
then $\partial_A F_A$ is a total derivative. 
Therefore we can represent $\mathcal{A}$ as a sum over critical points:
\begin{align}
	\mathcal{A}_{\Omega} &=\lim_{A\to\infty}\int_{\Gamma}F_A(\Omega)\nonumber\\
			  &=\sum_{t^m_*:\text{critical}} 
	\left\langle \prod_i^{n-3}(\mu_i(t^m_*),b)\prod_j^{n-3}(e_j(t^m_*),\tilde{b}) \;\frac{1}{\mathrm{det}\Phi}
	\;\Omega\right\rangle_{g(t^m_*)},\nonumber\\
	\Phi &:=(\Phi_{ij})_{ij}:=(\partial_{t_i}(e_j,\tilde{T}))_{ij}.
	\label{eq:A}
\end{align}
We have ignored some overall constants that come from orderings of fermionic things and Gaussian integrals.
The localization is a general feature of $\mathcal{A}$ defined by \eqref{eq:defA}
and does not depend on the choice of the theory $\mathcal{T}$ nor the choice of (covariant) fixing of 2d gauge symmetries. 

Further, the result \eqref{eq:A} can be formally rewritten as
\begin{align}
	\mathcal{A}_{\Omega}
	=
	\int_{\mathcal{M}} 
	\left\langle \prod_i^{n-3}(\mu_i(t),b)\mathrm{d}t_i\prod_j^{n-3}(e_j(t),\tilde{b}) \;\prod_{j}^{n-3}\bar{\delta}((e_j,\tilde{T}))
	\;\Omega\right\rangle_{g(t)},
	\label{eq:A2}
\end{align}
where $\bar{\delta}$ is the formal delta $(0,1)$-from defined by
\begin{align}
	\bar{\delta}(f(t)):=\bar{\partial}\frac{1}{f(t)}.
\end{align}
This formula is equivalent to what appeared in \cite{Adamo:2013tsa}.

\subsection{Integrated vertex operators}
\label{sec:intver}
So far we have been working with the formalism where the coordinates of vertices are fixed.
Here we describe the ``integrated vertex'' formalism which is often used in the literature.

Choose $\mu_i$ and $e_i$ so that $(\mu_i,\mathcal{O})=(e_i,\mathcal{O})=\frac{1}{2\pi\mathrm{i}}\oint_{\gamma_{i+3}} \mathcal{O}$
for quadratic differential local operator $\mathcal{O}$ where $\gamma_{i+3}$ is a contour around the $i+3$ rd vertex $\mathcal{V}_{i+3}$ whose matter part takes the form of \eqref{eq:flatvert}.
Then, the moduli parameters $t_i$ can be identifed with the positions $z_{i+3}$ of the vertex operators.
$b$ and $\tilde{b}$ insertions in \eqref{eq:A} merely deletes $c$ and $\tilde{c}$ of $\mathcal{V}_{i+3}$,
and \eqref{eq:A2} becomes
\begin{align}
	\mathcal{A}=\left\langle \prod_{i=1}^3 c\tilde{c}V^\text{m}_i(z_i) \prod_{i=4}^{n}\int\mathrm{d}z_i\bar{\delta}(\oint_{\gamma_i}\tilde{T}^\text{m})V^\text{m}(z_i) \right\rangle.
\end{align}
The term $\tilde{T}^\text{gh}$ in $\tilde{T}$ does not enter because of the ghost number conservation.
Therefore, we recognize 
\begin{align}
	\int\mathrm{d}z_i\bar{\delta}(\oint_{\gamma_i}\tilde{T}^\text{m})V^\text{m}(z_i)
	=
	\int\mathrm{d}z_i\bar{\delta}(k \cdot P) V^\text{m}(z_i)
\end{align}
as the ``integrated'' version of the fixed vertex operator $\mathcal{V}_i$,
which is exactly what was used in \cite{Mason:2013sva}.

\subsection{Loop integrands}\label{sec:loop}
For loop amplitudes, we cannot expect that the definition \eqref{eq:defA} works.
The ambitwistor string calculates amplitudes of a particle QFT or a gravity, which is not UV-finite
as opposed to the usual string.
Therefore what we can get is at most the integrand of the integration in terms of loop momenta $q_I$ as is investigated in \cite{Adamo:2013tsa,Casali:2014hfa}, whose existence is guaranteed when we have a space-time action. The whole amplitude should diverge when integrate over loop momenta.

Therefore, we should consider the loop-momenta-fixed amplitudes $\mathcal{A}(q)$
as is described in \cite{D'Hoker:1988ta,D'Hoker:1989ai} for the usual string.

Take a symplectic basis $(A_I,B_I)$, $I=1,\cdots g$ of the first homology of the genus $g$ Riemann surface.
Inserting the delta functions which fix the internal momenta,
we define the momenta-fixed version of $F$ by
\begin{align}
	F(\ell;\Omega;\mu,e|\delta \mu,\delta e)
	:=\left\langle e^{-(\delta\mu,b)} e^{-(e,\tilde{T})-(\delta e,\tilde{b})} \prod_I^g\prod_\mu \delta\left(\mathrm{i}\ell_I^\mu+\oint_{A_I}\frac{1}{2\pi\mathrm{i}}P^\mu\right) \Omega \right\rangle_{g}.
\end{align}
The delta functions kill the zero modes of $P$, therefore this $F(\ell)$ is expected to be finite.
The constraint from the delta functions for the zero modes of $P$ is solved by
\begin{align}
	P^\mu=2\pi\sum_I \ell_I^\mu\omega_I+\text{non-zero or singular modes},
	\label	{eq:momfix}
\end{align}
where $\omega_I$ are the smooth Abelian differentials on the Riemann surface which satisfy $\oint_{A_I}\omega_J=\delta_{IJ}$.

The integration cycle $\Gamma(\ell)$ is defined in the same way as in subsection \ref{sec:Intcycle}, 
but now operators are evaluated in $F(\ell)$ and therefore $\Gamma(\ell)$ depends on $\ell$.
The momenta-fixed amplitude is obtained by
\begin{align}
	\mathcal{A}_\Omega(\ell) := \int_{\Gamma(\ell)}F(\ell;\Omega).
\end{align}
The formal expression of $\mathcal{A}(\ell)$ similar to \eqref{eq:A2} is
\begin{align}
	&\mathcal{A}_{\Omega}(\ell)
	=\nonumber\\
	&\int_{\mathcal{M}} 
	\left\langle \prod_i^{\mathrm{dim}\mathcal{M}}(\mu_i(t^m),b)\mathrm{d}t_i\prod_j^{\mathrm{dim}\mathcal{M}}(e_j(t^m),\tilde{b}) \;\prod_{j}^{\mathrm{dim}\mathcal{M}}\bar{\delta}((e_j,\tilde{T}))
	\prod_{I,\mu} \delta\left(\mathrm{i}\ell_I^\mu+\oint_{A_I}\frac{1}{2\pi\mathrm{i}}P^\mu\right)
	\;\Omega\right\rangle_{g(t)}.
	\label{eq:A3}
\end{align}
Note that the insertion of the delta function $\delta(\mathrm{i}\ell_I+\oint_{A_I}\frac{1}{2\pi\mathrm{i}}P^\mu)$ explicitly breaks the invariance under the modular transformations which act on cycles $A_I$.
Therefore the solutions of the scattering equations, which depends on $\ell$, is not ensured to be in a single fundamental region when we move the external and loop momenta. 
It is not clear what thimbles $\mathcal{J}_m$ should be chosen as the summands of $\Gamma$,
in other word what solutions of scattering equations should be summed up,
and we postpone this problem till future investigations.

\subsection{A remark on the definition of $F$}
\label{sec:remark}
Before going to the fermionic case, let us clarify how the insertion $e^{(e,\tilde{T})}$ should be defined in the definition of $F$ \eqref{eq:Fdef}, which will be important in the section \ref{sec:prop}.
Note that $\tilde{T}=-\frac12 P^2 +\tilde{T}^\text{gh}$ only consists of free fields, therefore we can treat such insertions by expanding fields by their modes.

Here we concentrate on the matter part $\tilde{T}^\text{m}=-\frac12 P^2$
because the ghost part $\tilde{T}^\text{gh}$ does not contribute as long as we consider physical standard vertices.
Consider the case when the $X_\mu$ dependences of all the vertices $\mathcal{V}_i(z_i)$ are only the form of the kinematic factors $e^{\mathrm{i}k_i\cdot X}$ and the vertices do not depend on derivatives of $X^\mu$.
Then the equation of motion of $X_\mu$ force the $P^\mu$ to satisfy
\begin{align}
	\bar{\partial}P^\mu(z) = \mathrm{i} \sum_i k_i^\mu \delta(z-z_i).
\end{align}
To gather with \eqref{eq:momfix}, this equation fixes the $P_\mu$ to be
\begin{align}
	P^\mu = \hat{P}^\mu := 2\pi \left(\sum_{I=1}^g \ell_I^\mu \omega_I + \sum_{i=1}^nk^\mu_i\omega(z;z_i;z_0)\right),
\end{align}
where $\omega(z;z_i)$ is the Abelian differential of third kind 
defined by $\omega(z;z_i;z_0)= \frac1{2\pi\mathrm{i}}\partial_z\log\frac{E(z,z_i)}{E(z,z_0)}$,
where $E(z,w)$ is the prime form on $\Sigma$.
The $z_0$ dependence cancels out by the momentum conservation.
Therefore, we can replace $e^{(e,\tilde{T}^\text{m})}$ by $e^{(e,-\frac12 \hat{P}^2)}$.

For the later use, we rewrite this result as
\begin{align}
	\left\langle e^{(e,\tilde{T}^\text{m})} \Omega \right\rangle_{XP}
	&=
	e^{(e,-\frac12 \widehat{\tilde{T}}^\text{m})} \langle \Omega \rangle_{XP},\nonumber\\
	\widehat{\tilde{T}}^\text{m}&:=\frac{\langle\tilde{T}^\text{m}\Omega\rangle_{XP}}{\langle \Omega \rangle_{XP}}=-\frac12 \hat{P}^2,
	\label{eq:trick}
\end{align}
where $\Omega$ is the product of vertex insertions of the $XP$ CFT which do not contain derivatives of $X$.
We emphasize that this expression is valid only for the free $XP$ CFT and for the special type of the insertion $\Omega$. The standard vertex operator insertions are the case.
The correlator including derivatives of $X$ can be obtained by take derivatives 
of \eqref{eq:trick} in terms of insertion points.

\section{Amplitudes from a chiral CFT with fermionic currents}
\subsection{Algebras and the BRST charge}
Next we generalize the construction of the amplitudes of the previous section to a chiral CFT $\mathcal{T}$ with fermionic spin $\frac32$ currents.
We start from the data
\begin{align}
	(\mathcal{T},T^\text{m},\tilde{T}^\text{m},T^\text{m}_{F,a}) \quad (a=1,\cdots,N),
\end{align}
where $\mathcal{T},T^\text{m},$ and $\tilde{T}\text{m}$ are those of the previous section, and $T^\text{m}_{F,a}$ are the additional fermionic primaries.
We will see that the central charge $c^\text{m}$ of $\mathcal{T}$ is constrained to be
\begin{align}
	c^\text{m}=52-11N,
	\label{eq:cmconst}
\end{align}
where $N$ is the number of fermionic currents.
If we require $\mathcal{T}$ to be unitary, $N$ should be $1,2,3$ or $4$.

As the flat space part, we use the $X^\mu P_\mu$ CFT and $N$ copies of a set of worldsheet fermions $\psi^\mu_a,\,(a=1,\cdots,N)$ and set $T_{F,a}^\text{m} = \psi_a^\mu P_\mu$.
The $N=2$ theory is known to produce the tree level amplitudes of 10 dimensional type II supergravity.
$N=1$ theory with a current algebra $\hat{\mathfrak{g}}$ contains the Yang-Mills amplitude in its tree level amplitude as a leading trace part of the current algebra correlator. For $N=3,4$, we will briefly argue what happens in section~\ref{sec:conc}, but we will not go in detail.

The currents satisfy
\begin{align}
	T^\text{m}(z)T^\text{m}_{F,a}(0)&\sim \frac{3}{2z^2}T^\text{m}_{F,a}(0)+\frac{1}{z}\partial T^\text{m}_{F,a}(0)\nonumber\\
	T^\text{m}_{F,a}(z)T^\text{m}_{F,b}(0)&\sim C_a \frac{2\delta_{a,b}}{z}\tilde{T}^\text{m}(0)\nonumber\\
	\tilde{T}^\text{m}(z)T^\text{m}_{F,a}(0)&\sim 0
	\label{eq:TFOPE}
\end{align}
in addition to \eqref{eq:TtilTOPE}. 
The coefficient $C_a$ is equal to $1$ when $T_{F,a}=\psi_a\cdot P$, but if we use other $T_{F,a}$, $C_a$ can be zero (otherwise one can rescale $T_{F,a}$ to make $C_a$ to be equal to $1$) and the argument in this paper will not significantly depends on $C_a$.
Note that this algebra is \emph{not} the usual superconformal algebra since the OPE between fermionic currents generate $\tilde{T}$ and not the stress energy tensor $T$. 
The key difference from the usual ones when $N\ge2$ is the absence of currents which generates $R$-symmetries in the OPE's. 
Therefore we are allowed not to gauge the $R$-symmetry current and we do so.

To quantize the theory with these gauge symmetries, we introduce two copies of $bc$ system and $N$ copies of $\beta\gamma$ system as ghosts. 
The BRST current is
\begin{align}
	j_B= cT^\text{m}+cT_{\tilde{b}\tilde{c}}+\frac12 cT_{bc}+\tilde{c}\tilde{T}^\text{m}+\sum_a(cT_{\beta_a\gamma_a}+\gamma_aT^\text{m}_{F,a}-C_a \tilde{b}\gamma_a^2)+\frac32\partial^2 c,
\end{align}
where the residue of the squared OPE is
\begin{align}
	\mathrm{Res}\;j_B(z)j_B(0)=\frac{52-11N-c^\text{m}}{12}c\partial^3c
	+\partial \left(c\tilde{c}\tilde{T}^\text{m}+\sum_a (c\gamma_aT_{F,a}^\text{m}-C_a c\tilde{b}\gamma_a^2)\right).
\end{align}
Therefore the BRST charge $Q$ is nilpotent when $c^\text{m}=52-11N$.
The variations of $b$ and $\beta$ ghosts under this BRST transformation  gives
\begin{align}
	Q\cdot b&=T,\nonumber\\
	Q\cdot \tilde{b}&=\tilde{T}^\text{m}-(\partial\tilde{b})c-2\tilde{b}\partial c=:\tilde{T}^\text{m}+\tilde{T}^\text{gh}=:\tilde{T},\nonumber\\
	Q\cdot \beta_a &= T_{F,a}^\text{m}+c \partial\beta_a +\frac{3}{2}(\partial c)\beta_a-2C_a\tilde{b}\gamma_a
	=:T_{F,a}^\text{m}+T_{F,a}^\text{gh}=:T_{F,a}.
\end{align}
As seen, there also exits the ghost contribution to the fermionic currents $T_{F,a}$, and one can check the relations \eqref{eq:TtilTOPE} and \eqref{eq:TFOPE} hold without the superscript $\text{m}$.

As in the bosonic case, $T$ and $\tilde{T}$ couple to $\mu$ and $e$, respectively.
In the present case we also have fermionic currents $T_{F,a}$ which couple to fermionic ``gravitino'' backgrounds $\chi_a\in \Omega^{(0,1)}(\Sigma,\mathcal{R})$ where $\mathcal{R}$ is a dual spin bundle on the Riemann surface $\Sigma$ whose square $\mathcal{R}^2$ is isomorphic to the tangent bundle $K^{-1}$ of $\Sigma$. The coupling is described by
\begin{align}
	S_{\chi T_{F}}:=\frac{1}{2\pi}\int \chi T_F =:(\chi,T_F).
\end{align}
We also use the symbol $(\,,\,)$ for the pairing between the elements of $\Omega^{(0,1)}(\Sigma,\mathcal{R}^{1})$ and $\Gamma(\Sigma,\mathcal{R}^{-3})$.

\subsection{Vertex operators}

As in the usual superstring, we can choose the periodicity of each fermionic current $T_{F,a}$ around a vertex.
Therefore there are $2^N$ types of vertices, each is labeled by an $N$-tuple of NS or R.

\paragraph{All-NS sector}
First, we consider all-NS vertices.
We restrict the vertices $\mathcal{V}$ to satisfy
\begin{align}
	b_{n}\mathcal{V}=0\,(n\ge 0),\;\beta_{a,r}\mathcal{V}=0\,(r\ge0).
	\label{eq:dconfs}
\end{align}
We call them standard vertices again.
For the NS sector, $r$ take a value in $\mathbb{Z}+1/2$.
$Q$-closed $\mathcal{V}$ vanishes also by the action of $L_n,\tilde{L}_n,G_{a,r}=\oint\frac{\mathrm{d}z}{2\pi\mathrm{i}}z^{r+\frac12}T_{F,a}$ for $n,r$ not less than $0$.
A physical operator should have the ghost number 2, and in this paper we use the picture number $-1$ NS operators for that.
Such an operator takes the form of 
\begin{align}
	\mathcal{V}=c\tilde{c}\prod_a\delta(\gamma_a)V^\text{m}
\end{align}
where $V^\text{m}$ is a conformal primary if weight $2-\frac N2$ which vanishes when acted by $\tilde{L}_n,G_{a,r}$ for $n\ge 0,r\ge0$. The weight is determined by the fact that $\delta(\gamma)$ has weight $\frac12$.

For type II ambitwistor string, $V^\text{m}$ is taken to be
\begin{align}
	V^\text{m}=\psi_1\cdot\epsilon_1 \psi_2\cdot\epsilon_2 e^{\mathrm{i}k\cdot X},
\end{align}
which represents the space-time field with two indices.

For the $N=1$ ambitwistor string with a current algebra $\hat{\mathfrak{g}}$ as a internal CFT,
the vertex operator of a gluon is
\begin{align}
	V^\text{m}=\psi\cdot\epsilon\,j_a e^{\mathrm{i}k\cdot X},
\end{align}
and the leading trace part of correlators of $j_a$ gives the CHY formula for the pure Yang-Mills at tree level.

\paragraph{R sector}
For vertices including at least one R sector, we also impose the condition \eqref{eq:dconfs}, and the label of modes $r$ is in $\mathbb{Z}$.
We use the picture number $-1/2$ R sector operators, thus a physical operator looks like
\begin{align}
	\mathcal{V}=c\tilde{c}\prod_{a:\text{NS}}\delta(\gamma_a)\prod_{a:\text{R}}\Theta_a V^\text{m}
	\label{eq:physV2}
\end{align}
where $\Theta_a$ is the $\beta_a\gamma_a$ system spin operator whose weight is $3/8$.
$V^\text{m}$ should include corresponding R-sector vacuum operators, and should have conformal weight $2-\frac{N_{\text{NS}}}{2}-\frac{3N_{\text{R}}}{8}$ when $\mathcal{V}$ is a NS${}^{N_\text{NS}}$R${}^{N_\text{R}}$ operator.
It should also be annihilated by $\tilde{L}_n,(n\ge0)$ and $G_{a,r},(r\ge0)$.

As explained in \cite{Witten:2012bh}, it is more precise to treat R vertices in the way that $\beta\gamma$ system takes values in the line bundles which is not a tensor power of the spin bundle on Riemann surfaces instead of considering the ghost spin operators $\Theta$.
Affected by this, the corresponding fermionic current $T_{F,a}$ and gravitino background $\chi_a$ should take values in $\Gamma(\Sigma,\mathcal{R}^{-1}_a\otimes K)$ and $\Omega^{(0,1)}(\Sigma,\mathcal{R}_a)$ where 
$\mathcal{R}_a$ satisfies
\begin{align}
	\mathcal{R}^2_a\simeq K^{-1}\otimes \mathcal{O}(-D_\text{a,R})
	\label{eq:Rdef}
\end{align}
with the divisor $D_{a,\text{R}}:=\sum_{i:\text{R}} z_i$ of all R-punctures with respect to $T_{F,a}$.
For $\mathcal{R}_a$ to be well defined, the number of R vertices should be even for each $a$.

\paragraph{GSO projection}
As in the case of usual superstring, we should impose some GSO projection on the state space.

Let $F_a$ denote the worldsheet spinor number which counts $\psi_a^\mu$ and $\gamma_a$, and anticommutes with $T_{F,a}$ and commutes with $j_B$.
Then we project in terms of (some combinations of) $F_a$.

For example, to get type II ambitwistor string, we impose type II GSO projection on $N=2$ ambitwistor string as in the usual superstring case.
By projection in the NSNS sector, the ``tachyonic'' states $c\tilde{c}\delta(\gamma_1)\delta(\gamma_2)e^{\mathrm{i}k\cdot X}$ are projected out.
In the ambitwistor string case such ``tachyonic'' states are not BRST-closed from the beginning, but GSO projection on these states will be important to give the correct physical propagators of type II supergravity.

\subsection{Holomorphic top-form on a bundle over moduli}
As in the bosonic case, we define a holomorphic top-form on a bundle over the moduli $\mathcal{M}$.
Note that there is no room to the supermoduli of super Riemann surfaces to enter which is important to the usual superstring, because the algebra \eqref{eq:TFOPE} do never contain any superconformal algebras.
Instead, we will see the fiber direction of $T^*\mathcal{M}$ should be ``supersymmetrized''.

As a first step we work on the infinite dimensional supermanifold $\mathfrak{T}^*\mathscr{J}$ of ungauged (up to the Weyl transformation) backgrounds 
defined by
\begin{align}
	\mathfrak{T}^*\mathscr{J}:=\{(g,e|\chi_a)\}/\text{Weyl transf.}
\end{align}
This is an infinite dimensional $\mathbb{Z}_2$-graded vector bundle over $\mathscr{J}_\text{spin}$ including $T^*\mathscr{J}_\text{spin}$ where $\mathscr{J}_{\text{spin}}$ is the covering space of $\mathscr{J}$ whose covering corresponds to $2^{gN}$ choices of $\mathcal{R}_a$ which satisfies \eqref{eq:Rdef} for each $a=1,\cdots,N$.
We introduce fermionic auxiliary fields $\delta g$ and $\delta e$ and bosonic auxiliary fields $\delta \chi_a$, which couple to the corresponding $b$,$\tilde{b}$ and $\beta_a$ ghosts by the extended action
\begin{align}
	S_\text{ext}=(\delta \mu,b)+(\delta e,\tilde{b})-\sum_a (\delta \chi_a,\beta_a).
\end{align}
The BRST operator $Q$ is extended so that it acts on background fields by
\begin{align}
	[Q,\mu]=\delta \mu,\;[Q,e]=\delta e,\,\{Q,\chi_a\}=\delta\chi_a.
\end{align}
We recognize the auxiliary fields $\delta \mu,\delta e,\delta \chi_a$ as 1-forms on the supermanifold $\mathfrak{T}^*\mathscr{J}$.
For the definitions of the forms on supermanifolds, see \cite{Witten:2012bg}.

We define a form $F$ on $\mathfrak{T}^*\mathscr{J}$ by
\begin{align}
	F(\Omega;\mu,e,\chi_a|\delta \mu,\delta e,\delta \chi_a)
	:=\left\langle e^{-(\delta\mu,b)} e^{-(e,\tilde{T})-(\delta e,\tilde{b})-\sum_a(\chi_a,T_{F,a})+\sum_a(\delta \chi_a,\beta_a)}  \Omega \right\rangle_{g}.
	\label{eq:Fdeffer}
\end{align}
We have the same equation as in the bosonic case:
\begin{align}
	\mathrm{d}F(\Omega)=F(Q\cdot \Omega).
\end{align}

We would like to reduce the form $F$ to a form over the finite dimensional supermanifold 
\begin{align}
	\mathfrak{T}^*\mathcal{M}:=\mathfrak{T}^*\mathscr{J}/\mathcal{D}
\end{align}
which parametrises the background fields up to the group $\mathcal{D}$ of gauging.
$\mathcal{D}$ is generated by $(v,\alpha|\epsilon_a)$ where $v$ and $\alpha$ are in $\Gamma(\Sigma,T)$ 
and $\epsilon_a$ are fermionic sections of $\mathcal{R}_a$.
The bosonic generators $v,\alpha$ should vanish at all points where vertices sit,
and fermionic generators $\epsilon_a$ should vanish at points where NS vertices sit.
The (complex) dimension of the supermanifold $\mathfrak{T}^*\mathcal{M}$ is $6g-6+2n|(2g-2)N+N_\text{NS}+\frac{1}{2}N_\text{R}=:D_\text{even}|D_\text{odd}$
where $n$ is the number of vertices, $N_\text{NS}$ and $N_\text{R}$ are the total number of NS and R vertices, and $N$ is the number of fermionic currents.
For example, an NSNS vertex increases $N_\text{NS}$ by 2, and an NSR vertex increase both $N_\text{NS}$ and $N_\text{R}$ by 1.
There are deviations from this formula when $g$ and $n$ are small.

The orthogonality of $F$ with the vector fields generated by $\mathcal{D}$ is again ensured by the equations of motions of fermionic and bosonic ghost systems and the conditions of vertex operators \eqref{eq:dconfs}.
Therefore we can identify $F$ with a form on $\mathfrak{T}^*\mathcal{M}$.
We denote the local holomorphic coordinates of $\mathfrak{T}^*\mathcal{M}$ as $(t_i,a_i|\eta_{a,i})$,
and take a set of representatives $g(t),e(t,a),\chi_a(t|\eta_a)$ of background fields.
The dependence of $e$ and $\chi_a$ in $a$ and $\eta_a$ is taken to be linear as $e=\sum_j a_je_j$ and $\chi_a=\sum_k\eta_{a,k}\chi_{a,k}$.

If all of the vertices are physical, therefore look as \eqref{eq:physV2},
$F$ is a superdegree $D_\text{even}|D_\text{odd}$ form.
Namely, $F$ scales as $F\to \lambda^{D_\text{even}} F$ when all of $\mathrm{d}t_i$ and $\mathrm{d}a_i$ are scaled by $\lambda$,
and as $F\to \lambda^{-D_\text{odd}}F$ when all of $\mathrm{d}\eta_{a,i}$ are scaled by $\lambda$.
This scaling law for $\mathrm{d}\eta_{a,i}$ is explained by the rescaling of zero modes of $\beta$ ghost.

\subsection{Integration cycle and localization}
A superdegree $D_\text{even}|D_\text{odd}$ form over a ``real'' \footnote{
The more suitable terminology for the real structure of $\mathfrak{T}^*\mathcal{M}$ is a ``cs'' manifold. See\cite{Witten:2012bh}.} supermanifold of real dimension $2D_\text{even}|D_\text{odd}$ can be integrated over $D_\text{even}|D_\text{odd}$ dimensional sub-supermanifold.

Therefore we can use the same procedure as the bosonic case.
We define the integration cycle $\Gamma$ as we did in subsection~\ref{sec:Intcycle} 
using $h=-\Re((\mu(t),T)+(e(t,a),\tilde{T}))$ as the Morse function.

The amplitude is
\begin{align}
	\mathcal{A}_\Omega=\int_{\Gamma}F(\Omega)
\end{align}
again.
Although we should consider the loop-momenta-fixed version of this equation for $g\ge1$ as in the subsection \ref{sec:loop},  we suppress the notations related to the momenta fixing in the rest of this section.
The decoupling of BRST-exact states also holds.
The localization now gives
\begin{align}
	\mathcal{A}
		&=
	\sum_m \int
	\left\langle \prod_i^{D_\text{even}}(\mu_i(t^m),b)\prod_j^{D_\text{even}}(e_j(t^m),\tilde{b}) \;	
	e^{\sum_a \{Q,(\chi_a(t^m,\eta_a),\beta_a)\}}
	\frac{1}{\mathrm{det}\Phi}
	\;\Omega\right\rangle_{g(t^m)},\nonumber\\
	\Phi &:=(\Phi_{ij})_{ij}:=(\partial_{t_i}(e_j,\tilde{T}))_{ij}.
	\label{eq:AF}
\end{align}
The integration should be performed along the fermionic direction of $\mathfrak{T}^*\mathcal{M}$ restricted to each critical points $p_m\in T^*\mathcal{M}$.
The part related to the fermionic direction should be regarded as
\begin{align}
	\int	e^{\sum_a \{Q,(\chi_a(t^m,\eta_a),\beta_a)\}}
	&= \int \prod_a^N\prod_k^{D_\text{odd}} \delta(\mathrm{d}\eta_{a,k}) \delta((\chi_{a,k},\beta))e^{-\sum_{a,k}\eta_{a,k}(\chi_{a,k},T_{F,a})}+\cdots\nonumber\\
	&= \prod_a^N\prod_k^{D_\text{odd}}  \delta((\chi_{a,k},\beta))(\chi_{a,k},T_{F,a})+\cdots.
	\label{eq:chiins}
\end{align}
The ellipses represent terms aring from the regularization of the divergence caused by the collision of $\beta$ and $T_{F,a}$,
which can be determined from the BRST-invariance.
If the bases of the ``gravitino'' backgrounds $\chi_{a,k}$ are taken to be delta functions:
\begin{align}
	\chi_{a,k}(z)= \delta(z-w_{a,k}),
\end{align}
the insertion \eqref{eq:chiins} becomes a product of picture changing operators $X_a(w_{a,k})$ defined by
\begin{align}
	X_a(w) := T_{F,a}(w)\delta(\beta_a(w))-C_a \partial\tilde{b}(w)\delta'(\beta_a(w)),
\end{align}
where $C_a$ is the coefficient of OPE \eqref{eq:TFOPE},
and the amplitude is
\begin{align}
	\mathcal{A}=
	\sum_m 
	\left\langle \prod_i^{3g-3+n}(\mu_i(t^m),b)\prod_j^{3g-3+n}(e_j(t^m),\tilde{b}) \;	
	\prod_a\prod_k X_a(w_{a,k})
	\frac{1}{\mathrm{det}\Phi}
	\;\Omega\right\rangle_{g(t^m)}.\nonumber\\
\end{align}

When $g=0$ and all vertices are of type all-NS,
we are able to take $\mu_i$ and $e_i$ to be the same ones as in subsection \ref{sec:intver},
and to collide the $k$th picture changing operator  $X_a(w_{a,k})$ with the $k+2$th vertex $\mathcal{V}_{k+2}$.
Then the amplitude can be represented as
\begin{align}
	\mathcal{A}=\left\langle \prod_{i=1,2}\left( c\tilde{c}\prod_a\delta(\gamma_a)V_i^\text{m}(z_i) \right)
		c\tilde{c}\widehat{V}^\text{m}_3(z_3)
	\prod_{i=4}^{n}\int\mathrm{d}z_i\bar{\delta}(\oint_{\gamma_i}\tilde{T}^\text{m})\widehat{V}^\text{m}_i(z_i)
	\right\rangle,
\end{align}
where $\widehat{V}^\text{m}=\prod_a G_{a,-1/2}\cdot V^\text{m}$, which is the same formula as in \cite{Mason:2013sva}.

\section{Propagators and factorization}
\label{sec:prop}
In this section we investigate the space-time propagators and the factorization properties of the amplitudes defined in the previous sections.

An important property of the tree level scattering equations is that
when an internal momentum is on-shell, there are solutions on the corresponding boundary divisor of $\mathcal{M}_{g=0,n}$ \cite{Cachazo:2013gna}.
Using that fact, the factorization properties for the tree level CHY formulas are shown in the literature
\cite{Cachazo:2013hca,Cachazo:2013iea,Cachazo:2014xea}.
At the one-loop level, the property for the typeII ambitwistor string is investigated in \cite{Adamo:2013tsa,Casali:2014hfa} using the explicit expression of the one-loop scattering equations and the CFT correlators.

In the ambitwistor string setup, we expect that it is possible to re-prove these facts in the more abstract language of 2d CFTs, and generalize them to loop level.
For usual string, the factorization property is investigated in \cite{Cohen:1985sm} and is well reviewed in \cite{Witten:2012bh} and we follow the arguments there. For the factorization properties of the twistor string, see \cite{Adamo:2013tca}.
The arguments similar to what we are going to do can be found in \cite{Geyer:2014lca,Lipstein:2015rxa} for soft external or internal momenta.

\subsection{Bosonic case}

As is well known, an irreducible component of the boundary of the (Deligne-Munford compactification of the) moduli space $\mathcal{M}_{g,n}$ of genus $g$ and $n$-punctured Riemann surfaces is isomorphic to
$\mathcal{M}_{g_L,n_L}\times \mathcal{M}_{g_R,n_R}$ with $g_L+g_R=g,n_L+n_R=n+2,$ and $n_L,n_R\ge 1$
or $\mathcal{M}_{g-1,n+2}$. The former case is called a separating degeneration of Riemann surfaces, while the later is a non-separating degeneration.
We focus on the separating case, but the argument goes in the same way for the non-separating cases.

Nearly degenerated Riemann surfaces can be constructed by connecting punctures with ``pluming fixtures''.
Take a punctured Riemann surface $\Sigma_L\in\mathcal{M}_{g_L,n_L}$ and choose a local coordinate $\xi_L, |\xi_L|\le1$ around one of the punctures. Take another $\Sigma'\in\mathcal{M}_{g_R,n_R}$ and a local coordinate $\xi_R, |\xi_R|\le1$ around a puncture of $\Sigma'$.
Then we glue the two local coordinates $\xi_L$ and $\xi_R$ by imposing
\begin{align}
	\xi_L\xi_R = q.
	\label{eq:glue}
\end{align}
The obtained Riemann surface belongs to $\mathcal{M}_{g_L+g_R,n_L+n_R-2}=\mathcal{M}_{g,n}$ and near 
the boundary $D_{g_L,g_R,n_L,n_R}$, which is isomorphic to $\mathcal{M}_{g_L,n_L}\times\mathcal{M}_{g_R,n_R}$.
The gluing parameter $q$ plays the roll of the complex coordinate of $\mathcal{M}_{g_L+g_R,n_L+n_R-2}$ transverse to the boundary $D_{g_L,g_R,n_L,n_R}$, and $q=0$ corresponds to the points on the boundary.
The gluing condition \eqref{eq:glue} is compatible with $|\xi_L|,|\xi_R|\le1 $ when $|q|\le 1$
therefore $|q|\le1$ define a patch $\mathcal{U}$ of $\mathcal{M}_{g,n}$ around the boundary $D_{g_L,g_R,n_L,n_R}$.

The coordinate transformation $\rho=\log \xi_L$ maps the local patch around the chosen puncture of $\Sigma$ to a cylinder.
As the patch around the puncture of $\Sigma'$ is defined by $\xi_R\le1$,
$\xi_L$ can take $|q|\le |\xi_L| \le 1$ and the cylinder represented by $\rho$ has the length $s:=-\log |q|$, and is twisted by the angle $\phi:=\mathrm{arg}q$.

Let us use $S:=-\log q=s+\mathrm{i}\phi$ as the coordinate of the moduli of the cylinder instead of $q$.
The corresponding Beltrami differential $\mu_S=g^{\rho\bar\rho}\partial_Sg_{\bar\rho\bar\rho}$ is equal to $\frac1s$ in the $\rho$ coordinate of the cylinder.
Thus, the pairing between $\mu_S$ and the $b$ ghost is calculated to be
\begin{equation}
	(\mu_S,b) = \int_{|\rho|=0}^{|\rho|=s} \mu_S b = \oint_{\rho=0}^{\rho=2\pi\mathrm{i}} b =b_0.
\end{equation}
For the Lagrange multiplier Beltrami differential background $e$, we use the same the basis $\mu_S$
and expand it as $e=\tilde{s}\mu_S$.

In $\mathcal{U}$, the form $F$ on $T^*\mathcal{M}$ can be written as 
\begin{align}
	F= \sum_{\Psi_L,\Psi_R\in\mathcal{H}} 
	F_L(\Omega'_L \Psi_L(\xi_L=0))
	\braket {\Psi_L | e^{-\{Q,sb_0+\tilde{s}b_0\}} |\Psi_R}
	F_R(\Psi_R(\xi_R=0) \Omega'_R ),
	\label{eq:propsum}
\end{align}
where $\Omega'_{L,R}$ are the external vertices on $\Sigma_{L,R}$,
$F_{L,R}$ are the form on $T^*\mathcal{M}_{L,R}$ defined by $\Omega'_{L,R}$ and internal states $\Psi_{L,R}$,
and the sum is taken over a pair of bases $\Psi_{L,R}$ of the total Hilbert space $\mathcal{H}$ of string states.
By the ghost number conservation, both $\Psi_L$ and $\Psi_R$ can be assumed to have $bc$ ghost number $N_{bc}$ and $\tilde{b}\tilde{c}$ ghost number $N_{\tilde{b}\tilde{c}}$ equal to $1$ and we can ignore the ghost part of $\tilde{L}_0$.
\footnote{The operator $\tilde{T}^\text{gh}$ mixes two ghost systems,
but the coupling $(e,\tilde{T}^\text{gh})$ cannot contribute when we use standard on-shell vertices as external vertices as mentioned before.}
We concentrate on the 26 dimensional flat bosonic ambitwistor case at first, and will mention on the case of non-trivial internal CFTs later.
For the flat bosonic ambitwistor string, the weight $h$ of the states is greater than or equal to $-2$.

To compress the expressions we define the symbols $\llangle \rrangle_{h}$ and $\llangle \rrangle^S$ by
\begin{align}
	\llangle \mathcal{O} \rrangle_h &:= 
	\sum_{\Psi_L,\Psi_R\in\mathcal{H}} 
	F_L(\Omega'_L \Psi_L(\xi_L=0))
	\braket {\Psi_L | b_0 \tilde{b}_0 \delta(L_0-h)\mathcal{O} |\Psi_R}
	F_R(\Psi_R(\xi_R=0) \Omega'_R ),\nonumber\\
	\llangle \mathcal{O} \rrangle^S &:= 
	\sum_{\Psi_L,\Psi_R\in\mathcal{H}} 
	F_L(\Omega'_L \Psi_L(\xi_L=0))
	\braket {\Psi_L | b_0\tilde{b}_0 \mathcal{O} e^{-s L_0}|\Psi_R}
	F_R(\Psi_R(\xi_R=0) \Omega'_R ),\nonumber\\
	&=\sum_h \llangle \mathcal{O} \rrangle_h q^{h}.
\end{align}
where the operator $\delta(L_0-h)$ projects the states to have $L_0$ eigenvalue $h$.

By \eqref{eq:trick},
$F$ can be rewritten as
\begin{align}
	F= \mathrm{d}s\mathrm{d}\tilde{s}e^{-\tilde{s}\llangle \tilde{L}_0 \rrangle^S/\llangle 1 \rrangle^S}\llangle 1 \rrangle^S,
\end{align}
and the saddle point equation is
\begin{align}
	\llangle \tilde{L}_0 \rrangle^S/\llangle 1 \rrangle^S=0.
	\label{eq:saddle}
\end{align}
Again we emphasize that \eqref{eq:trick} uses the special property of the $XP$ CFT which is free.

It is convenient to split the operator $\tilde{L}_0$ as $\tilde{L}_0= k^2 +\tilde{L}_0'$.
In the loop-momenta-fixed amplitudes, the internal momentum $k$ which goes through the cylinder is always fixed, therefore $k^2$ can be treated as a number.
The remaining part $\tilde{L}_0'$ does not depend on $k$, and acts on, for example, $\partial X$ as
$\tilde{L}_0'\cdot \partial X_\mu = P_\mu$.
Then the equation~\eqref{eq:saddle} becomes
\begin{align}
	\llangle \tilde{L}_0'\rrangle^S/\llangle 1 \rrangle^S = -k^2.
	\label{eq:saddle2}
\end{align}

$\tilde{L}_0'$ is zero when restricted to the space of states with $L_0=-2$ which contains only $c\tilde{c}e^{\mathrm{i}k\cdot X}$.
Therefore, $\llangle \tilde{L}_0 \rrangle_s/\llangle 1 \rrangle_s = \mathcal{O}(q^1) $ and, for sufficiently small $k^2$, there exists a solution $q_*$ which satisfies
\begin{align}
	q_* = -k^2\frac{\llangle 1\rrangle_{-2}}{\llangle \tilde{L}_0 \rrangle_{-1}}+\mathcal{O}((k^2)^2).
	\label{eq:qstar}
\end{align}
This solution $q_*$ approaches $0$ when the internal momentum $k$ get close to on-shell.
In other words, when internal momentum is on-shell, there exist solutions of scattering equations
on the corresponding boundary divisor in the moduli $\mathcal{M}$.
This phenomenon is already investigated for tree scattering equations \cite{Cachazo:2013gna} and for one-loop scattering equations \cite{Adamo:2013tsa}.

The Gaussian integration near the solution $q_*=e^{-S_*}$ along the Lefschetz thimble gives
\begin{align}
	\frac{\llangle 1 \rrangle^{S_*}}{\partial_{S} \left(\llangle \tilde{L}_0'\rrangle^{S}/\llangle 1 \rrangle^{S}\middle)\right|_{S=S_*} }
	&=
	-\frac{\llangle 1 \rrangle^{S_*}}{\llangle L_0 \tilde{L}_0'\rrangle^{S_*}/\llangle 1 \rrangle^{S_*}+k^2\llangle L_0 \rrangle^{S_*}/\llangle1\rrangle^{S_*} }
	\nonumber\\
	&= 
	-\frac{\llangle 1 \rrangle^{S_*}}{q_*\llangle L_0 \tilde{L}_0'\rrangle_{-1}/\llangle 1 \rrangle_{-2}+k^2\llangle L_0 \rrangle_{-2}/\llangle1\rrangle_{-2}+\mathcal{O}((k^2)^2) }\nonumber\\
	&= 
	\frac{\llangle 1 \rrangle_{-2}q_*^{-2}+\llangle 1 \rrangle_{-1}q_*^{-1}+\llangle 1 \rrangle_{0}}{k^2}+\mathcal{O}((k^2)^0).
	\label{eq:bosprop}
\end{align}

This result is not what is expected from physically meaningful theories.
The first and the second terms give unpleasant $\mathcal{O}((k^2)^{-3})$ and $\mathcal{O}((k^2)^{-2})$ contribution. Those come from the operators of conformal weight $-2$ and $-1$, which cannot be BRST invariant.
The $\mathcal{O}((k^2)^{-1})$ part also contains unpleasant contributions from unphysical operators like $P_\mu\partial_\nu X e^{\mathrm{i}k\cdot X}$ and higher order terms of $q^{-2}_*,q^{-1}_*$.
Therefore, the gravity part of the bosonic ambitwistor string looks pathological.

Let us comment on the case with 2 copies of internal current algebras as internal CFTs.
As mentioned and originally found in \cite{Mason:2013sva},
its double leading trace part reproduces the amplitudes of scalars with the interaction like
$f_{abc}\tilde{f}_{\tilde{a}\tilde b\tilde c}\phi^{a\tilde a}\phi^{b\tilde b}\phi^{c\tilde c}$ considered in \cite{Cachazo:2013hca}.
The full propagator of such ambitwistor string should contain dangerous terms like in \eqref{eq:bosprop}.

However, the problem does not exist in the double leading trace part.
The states which contribute to the double leading trace part should contain
the current operators $j$ and $\tilde{\jmath}^{\tilde a}$ of both current algebras.
The only such operator which contribute to the propagator is $c\tilde{c}j^a\tilde{\jmath}^{\tilde a}e^{\mathrm{i}k\cdot X}$ which is physical when $k^2=0$.
The denominator of \eqref{eq:bosprop} is not changed since it is kinematically determined,
therefore the propagator is physical when restricted to the double leading trace part,
which is consistent with the analysis of \cite{Cachazo:2013iea}.

\subsection{Fermionic case}
The calculation is almost the same when there exist fermionic spin $3/2$ currents $T_F$.
The sum in \eqref{eq:propsum} should be taken in terms of picture $-1$ NS states and $-1/2$ R states.
We also change the definitions of the symbol $\llangle \rrangle_h$ so that they take the sum over only states with such picture numbers.
We again have the solution \eqref{eq:qstar} since scattering equations are kinematically determined,
and have the same denominator $k^2$ in the propagator.

For all-NS sector, as there exists no zero-mode of gravitino backgrounds on the cylinder with NS-boundary conditions, the numerator is the same except for the shift of level due to $\delta(\gamma)$ operators and the insertion of the GSO-projection operator $\Pi_\text{GSO}$ which comes from the spin structure sum.
If the internal CFT is unitary, the resulting propagator is 
\begin{align}
	\frac{\sum_{h=-2+N/2}^{0} \llangle \Pi_\text{GSO} \rrangle_h q_*^h}{k^2}+\mathcal{O}((k^2)^0),
\end{align}
where $N$ is the number of the fermionic currents.
To have a healthy propagator, the projection $\Pi_\text{GSO}$ should kill all the unphysical states with standard ghost  contribution $c\tilde{c}\prod_{a}\delta(\gamma_a)$ and weight not more than 0.

For $N=1$ theory with a current algebra as an internal CFT, $\Pi_\text{GSO}$ is too small to eliminate all problematic states.
Instead, we again concentrate on the leading trace part and get tree level Yang-Mills amplitudes as obtained in \cite{Mason:2013sva}.

For $N=2$, the GSO projection associated to two fermionic currents kills all the problematic states, if we choose type II projection. Therefore, the type II ambitwistor string has correct NS-sector propagator.
For $N=3$, we again are able to choose $\Pi_\text{GSO}$ so that it kills unwanted states, and finally unpleasant states are absent from the beginning for $N=4$.

A cylinder with R-boundary conditions has gravitino zero modes and corresponding $\beta_a$ ghost zero modes. 
Therefore, we should include additional $e^{\{Q,\eta_a \beta_a\}}$ contribution in \eqref{eq:propsum}
where $\eta_a$ is the corresponding odd modulus.
Integrating $\eta_a$, the propagator for NS${}^A$R${}^B$ sector becomes
\begin{align}
	\frac{\sum_{h=-2+A/2+3B/8}^{0} \llangle \prod_{a\in\text{R}}\delta(\beta_{a,0})G_{a,0}\Pi_\text{GSO} \rrangle_h q_*^h}{k^2}
	+\mathcal{O}((k^2)^0).
\end{align}
Again, one can check the absence of unpleasant terms from type II ambitwistor string.

\section{Conclusions and discussions}
\label{sec:conc}
\subsection{Conclusions}
In this paper, we performed the abstract investigation on the structures of the ambitwistor string from the worldsheet point of view.
We constructed the conformally invariant formulation of the amplitudes of the theory,
and made some analysis on its propagators.
\paragraph{Worldsheet aspects of ambitwistor string}
The ambitwistor string and its standard operators define 
a holomoriphic top-form $F$ on the cotangent bundle $T^*\mathcal{M}$ of
the moduli $\mathcal{M}$ of punctured Riemann surfaces or on its ``supersymmetrized'' version $\mathfrak{T}^*\mathcal{M}$, which is independent of the gauge fixing conditions.
To define the amplitude, a middle dimensional cycle in $T^*\mathcal{M}$ or $\mathfrak{T}^*\mathcal{M}$ should be chosen, and we adopted a cycle $\Gamma$ defined by the Morse flow.
Eventually the amplitude localizes on the solutions of scattering equations
and reproduces the formulas in \cite{Mason:2013sva,Adamo:2013tsa} in the case of genus $0$ and $1$.

\paragraph{Propagators}
The factorization properties of the amplitudes defined above are investigated.
We found that the Jacobian factor of the scattering equations give the usual $(k^2)^{-1}$ factor of the proper massless propagators, however we also found unphysical factors.
Nevertheless, such unpleasant factors are absent in the leading trace part of the internal current algebra, and in type II ambitwistor string.

\subsection{Future directions}
\paragraph{Search for meaningful internal systems}
This paper ensures that, in principle, ambitwistor string can have various internal chiral compact CFTs.
It should be interesting to consider how larger classes amplitudes of QFTs can be constructed as ambitwistor string theories.
For the theory to be physically meaningful, one should ensure the propagator is free from the contribution from unphysical vertex operators.

\paragraph{$N=3$ ambitwistor string and the DBI theory}
We proposed the possibility to consider the ambitwistor string with $N=3$ or $4$ fermionic currents.
The amplitudes of the $N=3$ theory should contain three factors of the reduced Pfaffians of matrices depending on momenta and polarizations in its summand.
Such summand is proposed in \cite{Cachazo:2014xea}
for the DBI theory.

Actually, one can quickly check that the NS${}^3$ sector of the flat $N=3$ ambitwistor string reproduces
the BI part of the formula.
A representative of a BRST cohomology which represents a space-time vector is
\begin{align}
	\mathcal{V}_{A}= c\tilde{c}\prod_{a=1}^3 \delta(\gamma_a) \epsilon\cdot \psi_1 e^{\mathrm{i}k\cdot X},
\end{align}
and its picture-raised form is 
\begin{align}
	\widehat{\mathcal{V}}_A= c\tilde{c}(P\cdot \epsilon+\psi_1\cdot\epsilon\psi_1\cdot k)\,k\cdot \psi_2\,k\cdot\psi_3 \,e^{\mathrm{i}k\cdot X}.
\end{align}
The three sets of fermions $\psi_{1,\mu},\psi_{2,\mu},\psi_{3,\mu}$ generate
three of reduced Pfaffians,
and the one produced by the $\psi_1$ correlator depends on both polarizations $\epsilon$ and momenta $k$ of external states,
while the others only contain momenta. 
The resulting tree amplitude is exactly what is proposed in \cite{Cachazo:2014xea} for the BI theory.

If we use other sets of fermions $\psi^\text{int}_i$ as the internal system,
we also have on-shell operators
\begin{align}
	\mathcal{V}_{\phi_i} = c\tilde{c}\prod_{a=1}^3 \delta(\gamma_a) \psi_i^\text{int} e^{\mathrm{i}k\cdot X},
\end{align}
which represent $i$th space-time scalar $\phi_i$.
The internal fermions produce (non-reduced) Pfaffians and again reproduce the result of the reference 
for DBI.

To make the total worldsheet theory anomaly-free, the dimension $D$ of space-time and the number of space-time scalars $N_{\phi}$ should satisfy $7 D+N_\phi = 38$. The possible combinations are
$(D,N_\phi)=(5,3),(4,10),(3,17),(2,24)$.
The meaning of these sets, possible ways of GSO projections and Ramond sectors are left to be researched.

\paragraph{$N=4$  ambitwistor string and Galileon theory with special coupling}
As above, tree amplitudes of $N=4 $ ambitwistor string should contain four of reduced Pfaffians.
Such amplitudes can be found again in \cite{Cachazo:2014xea}, which is for the Galileon theory with special values of the couplings so that the theory gets enhanced symmetry as stated in \cite{Hinterbichler:2015pqa}.

The on-shell operator of $N=4$ ambitwistor string is
\begin{align}
	c\tilde{c}\prod_{a=1}^4 \delta(\gamma_a) e^{\mathrm{i}k\cdot X}.
\end{align}
The tree level amplitude contains the quartic power of the reduced Pfaffian of a matrix consisting of momenta,
which again agree with the results given in \cite{Cachazo:2014xea}.

The dimension should be $D=2$ to achieve anomaly cancellation.
The propagator is physical for $N=4$ theory because there are no undesired states in $L_0\le 0$.
The Ramond sectors and supersymmetries are remain to be investigated, again.

\paragraph{Rationality of Loop integrands}
The loop integrands discussed in subsection~\ref{sec:loop} should be rational functions in terms of the external and internal momenta which is manifest in the Feynmann diagram calculations.
In the IR limit, this problem is discussed in \cite{Casali:2014hfa,Lipstein:2015rxa}
As stated in there, how to take the sum in terms of the solutions of loop-level scattering equations
is not obvious so far, and the way should be fixed so that the integrands achieve the rationality.

\paragraph{Relations to usual string theory}
In this paper, we just made a consistent procedure to define amplitudes by hand for ambitwistor string.
However, as is done at the level of the action in \cite{Mason:2013sva}, 
the whole procedure can possibly be deduced from infinite tension limit of the usual string theory
and may reveal the so-far-obscure relation between the scattering equations and the saddle point method of \cite{Gross:1987kza} along the way of \cite{Bjerrum-Bohr:2014qwa,Tourkine:2013rda,Caputa:2011zk,Caputa:2012pi}.

\section*{Acknowledgements}
The author sincerely thanks F. Cachazo and Y. Tachikawa for helpful discussions.
The author is partially supported by the Programs for Leading Graduate Schools, MEXT, Japan,
via the Advanced Leading Graduate Course for Photon Science.
The author is also supported by JSPS Research Fellowship for Young Scientists.
 
\appendix	
\section{No-ghost theorem for bosonic ambitwistor string}\label{sec:noghost}
In this appendix we show the no-ghost theorem for the bosonic ambitwistor string and 
that one can always choose a vertex operator $\mathcal{V}$ which is annihilated by $b_n$ and $\tilde{b}_n$ for $n\ge0$ as a representative of a $Q$-cohomology class.
The discussion is imported from section~4.4 of \cite{Polchinski:1998rq} in the case of the usual string and almost the same.
The ``No-ghost'' theorem here only ensures that there is no ghost in the BRST-cohomologies,
thus this theorem does not contradict with the unphysical form of propagators found in section~\ref{sec:prop}.

We assume that the CFT $\mathcal{T}$ contains the flat $P^\mu X_\mu$ CFT for $\mu=0,1$ with Lorentzian metric $\eta_{\mu\nu}=\mathrm{diag}(-1,1)$. 
A bit surprisingly, the argument for the usual string can be modified in spite of the difference of the theory of the light cone direction.
We work on the light cone coordinates $\eta_{+-}=\eta_{-+}=-1,\eta_{++}=\eta_{--}=0$, and expand fields as
\begin{align}
	X^\mu(z)=\sum_m\frac{X^\mu_m}{z^m},\qquad
	P^\mu(z)=\sum_m\frac{P^\mu_m}{z^{m+1}},
\end{align}
whose modes satisfy
\begin{align}
	[X^\mu_m,P^\nu_n]=\eta^{\mu\nu}\delta_{m+n,0}.
\end{align}
Then we define a number operator
\begin{align}	
	N^\text{lc}=\sum_{\substack{m\neq0\\m=-\infty}}^{m=\infty} (P^+_{-m}X^-_m - P^-_{-m} X^+_m).
\end{align}
Note that this operator is almost the space-time Lorentz generator, but we have removed the zero mode from that.
We consider states with nonzero momentum and go to a frame with $k^+\neq 0$ and impose $b_0=\tilde{b}_0=0$.

The BRST operator $Q$ splits into three parts
\begin{align}
	Q=Q_1+Q_0+Q_{-1},
\end{align}
where $Q_i$ has charge $i$ under $N^\text{lc}$.
The nilpotency of $Q$ implies that of $Q_1$.
The concrete form of $Q_1$ is
\begin{align}
	Q_1=k^+ \sum_{m\neq0}\left( -m c_{-m} X^-_m - \tilde c_{-m}P^-_m\right).
\end{align}
Then we define operators
\begin{align}
	R=\frac{1}{k^+}\sum_{m\neq0}\left(b_{m}P^+_{-m} + m\tilde{b}_{m}X^+_{-m}\right)
\end{align}
and
\begin{align}
	S:=&\{Q_1,R\}\nonumber\\
	=& \sum_{m\ge1} m\left(c_{-m}b_m+b_{-m}c_m
		+\tilde{c}_{-m}\tilde{b}_m+\tilde{b}_{-m}\tilde{c}_m \right.\nonumber\\
	&\hspace{2.6cm}\left.+X^-_{-m}P^+_{m}-P^+_{-m}X^-_m+X^+_{-m}P^-_m-P^-_{-m}X^+_m\right).
\end{align}

This operator $S$ commutes with $Q_1$, therefore the remaining argument is exactly the same as that of \cite{Polchinski:1998rq}.
Here we briefly rephrase that for completeness.

If $\ket{\psi}$ has positive $S$-number and $Q_1$-closed, then $\ket{\psi}=Q_1 S^{-1}R\ket{\psi}$ is $Q_1$ exact.
If $\ket{\psi}$ has $S$-number 0, then $\ket{\psi}$ is $Q_1$-closed because $0=Q_1S\ket{\psi}=SQ_1\ket{\psi}$ and $Q_1\ket{\psi}$ consists of positive $S$-number states.
Thus we have got
\begin{align}
	Q_1 \text{-cohomology} \cong \mathrm{Ker} \, S,
	\label{eq:Q1kerS}
\end{align}
which is the no-ghost theorem for $Q_1$. (Here $\cong$ means we can choose representatives of each elements of the left hand side from the right hand side.)

To relate this space to the $Q_B$-cohomology, we use $U=\{Q_0+Q_{-1},R\}$.
Note that $\Phi=\frac{1}{1+S^{-1}U}=\sum_{n\ge0} (-S^{-1}U)^n$ is a well-defined operator since $S^{-1}$ always acts on the states with $N^\text{nl}<0$ where $S$ is invertible. Since $S+U=S(1+S^{-1}U)$, $\Phi$ defines an isomorphism
\begin{align}
	\Phi:\mathrm{Ker} \, S \xrightarrow{\sim} \mathrm{Ker}\, (S+U).
\end{align}
The same argument as for \eqref{eq:Q1kerS} shows
\begin{align}
	Q_B \text{-cohomology} \cong \mathrm{Ker} \, (S+U),
	\label{eq:QBkerS+U}
\end{align}
therefore the proof is done.

Finally, we would like to show that there is always at least one state in a $Q$-cohomology class which is killed by $b_n$ and $\tilde{b}_n$ for $n\ge0$. It is enough to show that states in $\mathrm{Ker}\,(S+U)$ satisfy this property. 
We define a number
\begin{align}
	N'=
	 \sum_{m\ge1} \left(c_{-m}b_m+b_{-m}c_m
		+\tilde{c}_{-m}\tilde{b}_m+\tilde{b}_{-m}\tilde{c}_m
	+2X^-_{-m}P^+_{m} - 2 P^-_{-m}X^+_m\right).
\end{align}
This number counts various operators as
\begin{align}
	[N',b_n]&=-\sign(n)b_n,& [N',c_n]&=-\sign(n)c_n,\nonumber\\
	[N',X^-_{-l}]&=2X^-_{-l},& [N',P^+_{l}]&= -2P^+_l \;\;(l\ge1),
\end{align}
and so on. One can check $R$ has $N'$-number $-1$, and $Q_0+Q_{-1}$ does not contain terms with $N'$-number greater than or equal to 1. Therefore $\Phi \ket{\psi}$ should contain terms with $N'\le0$ for $\psi\in\mathrm{Ker}\,S$, but $N'$ should be no less than 0 for states, concluding $N'\Phi\ket{\psi}=0$. 
This statement includes what we wanted to show.

\bibliographystyle{utphys}
\bibliography{refs}
\end{document}